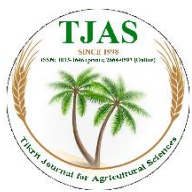
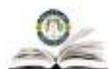

**Anwar Mohammed Raouf** *
**Kocher Omer Salih**
**Aram Akram Mohammad**

*Horticulture Department, College of Agricultural Engineering Sciences, University of Sulaimani, Kurdistan Region-Iraq*

# Examination of Some Nut Traits and Release From Dormancy Along With Germination Capacity in Some Bitter Almond Genotypes

## ABSTRACT

This study was conducted at College of Agricultural Engineering Sciences, University of Sulaimani, Kurdistan Region-Iraq so as to investigate some nut traits in 10 bitter almond genotypes, capacity of them to release from dormancy and finally germination ability. Nut traits were calculated, and stratified in a sand medium at 6±1°C in a refrigerator for 55 days, then they were sown in fine sand on August 22, 2021 for 29 days to calculate germination percentage. There were great discrepancies among genotypes in nut traits. Nut length was between (23.66-32.73 mm), nut width (18.77-21.84 mm), nut size (3.16-4.26 cm3), nut weight (2.67-4.13 g), kernel weight (0.6-0.99 g), shell weight (1.94-3.27 g) and shell thickness (2.31-3.37 mm). The results of release from dormancy and calculation of germination percentage trials showed that the highest nut numbers from G6 (100%), G4 (86.11%) and G2 (65.22%) were released from dormancy and the same genotypes gave the best germination percentage, particularly G6 and G2 both gave (73.33%) germination. Depending on the results of release percentage from dormancy and germination percentage, G6 and G2 along with G4 were the best genotypes.

## INTRODUCTION

Almond [*Prunus dulcis* (Mill.) D.A.Webb; syn. *P. amygdalus* Batsch] is a deciduous nut from stone fruit group. It is attributed to Rosaceae, subfamily Prunoideae. The wild species of almond found in Tian Shan mountain in western China, mountainous areas of Kurdistan, Turkestan, Afghanistan, Iran, and Iraq (Martínez-Gómez *et al*., 2007). Almond is cultivated in areas have subtropical Mediterranean climate with mild wet winter and warm, dry summer (Sorkheh *et al*., 2018). It is well-known for tolerance against harsh conditions and endurable along all stages of growth (Palasciano *et al*., 2014). It has mechanisms of both avoidance and to some degree osmotic adjustment as it is subjected to water stress periods (Yadollahi *et al*., 2011). Also, almond is used as rootstocks for budding and grafting stone fruits in stressful environments, particularly 'Bitter' genotypes (Rahemi and Yadollahi, 2005). The rootstocks from bitter almond are very tolerable to drought and the pests in soil compared to the ones from sweet almond (Parvaneh *et al.,* 2011). Although, the nuts of bitter almond are used to extract the essential oil which is important for cosmetic and pharmaceutical industries (Loghavi *et al*., 2011; Nasirahmadi and Ashtiani, 2017).

---

* *Corresponding author: E-mail*: aram.hamarashed@univsul.edu.iq

Bitter almond is economically processed to production bioenergy, and activated coal is achieved from nut shell of bitter almond as well (Trachi *et al*., 2014; Akubude and Nwaigwe, 2016).

Almond has been traditionally propagated via seeds, this has allowed production a large number of different genotypes which exhibit diverse flowering time, resistance to drought, resistance to pest and disease, seed traits, seed dormancy, and seed germination (Imani and Nagoya, 2006; García-Gusano *et al*., 2011). Seed quality influences the vigor of the seeds to produce new plants and the subsequent growth of them, and the quality of seeds is subjected to many ecological and genetic factors (Borji *et al*., 2007). Almond genotypes were exhibited different nut quality in terms of nut weight, kernel weight, double kernels, shell hardness, kernel shape, etc (Khadivi *et al*., 2019). Additionally, Sakar *et al*. (2019) found that nut and kernel length and sphericity of almond cultivars are under genetic control, but nut and kernel thickness was under environmental effect.

Besides, outer covering of seed (endocarp) and hormonal or physiological status exist in embryo induce seed dormancy in almond (García-Gusano *et al*., 2009). Stratification is the common method applied to the seeds to overcome seed dormancy in the dormant seeds. For almond seeds, it is observed that cold stratification is decisive to break seed dormancy, while almond cultivars differently responded to periods of stratification (García-Gusano *et al*., 2004). Therefore, this study aimed to evaluate breaking seed dormancy and germination along with some seed traits in some genotypes of bitter almond.

**MATERIALS AND METHODS**

The fruits of 10 bitter almond genotypes were collected in August 2020 from Barznja province, Sulaimani city, Kurdistan Region-Iraq (latitude 35º27'69" Nº 45º42'21"Wº, altitude 1154). The hull of the almond fruits was removed and the nuts dried in shade at room temperature for 5 days, then they stored at room temperature in paper envelop at College of Agricultural Engineering Sciences, University of Sulaimani, Kurdistan Region-Iraq until the time of using. The genotypes from 1 to 10 were marked as (G1, G2, G3, G4, G5, G6, G7, G8, G9 and G10). The picture of the nuts of the genotypes and their kernels depicted in figure (1).

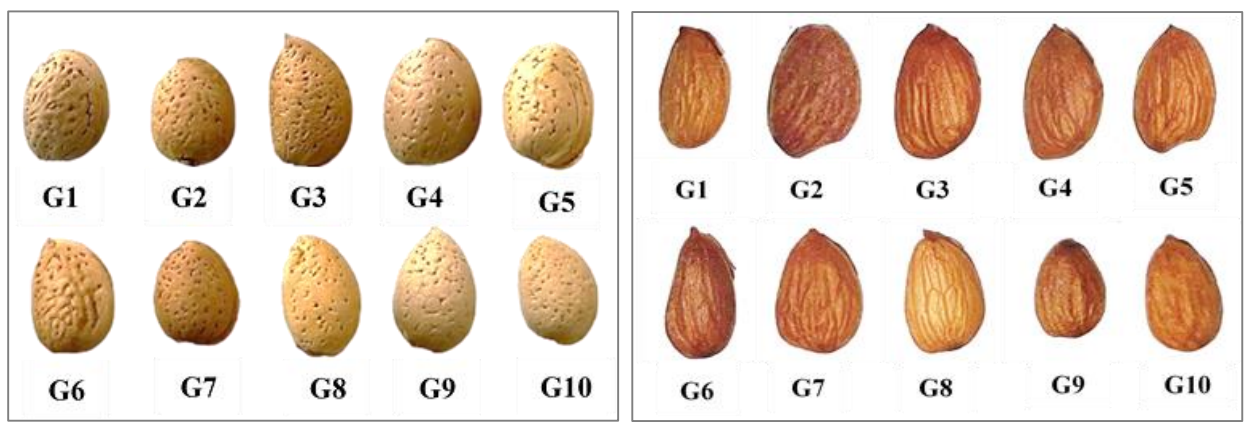

Figure (1) Picture of nut and kernel of the 10 bitter almond genotypes

**Nut trait measurements**

Nine nuts from each genotypes were selected to measure nut length and nut width used electronic digital caliper, nut size via submerging the nuts in water in a cylinder then determined the raised water level, nut weight, kernel weight, nut shell weight, nut shell thickness used electronic digital caliper which taken from middle of the nut shell. The data of nut traits analyzed used RCBD layout and 5% Duncan's Multiple Range Test for comparison of the means.

**Estimation of percentage of released nuts from dormancy**

The nuts of the bitter almond genotypes were stratified directly in a sand medium and placed in a refrigerator at 6±1°C on June, 28 2021, without application of any pre-treatment or disinfectant to the stratified nuts. The stratified nuts in containers laid out in RCBD with three replications inside the refrigerator, and in each replication 12 nuts were stratified, totally for the same genotype 36 nuts were sown in the three replications. On August, 22 2021, after 55 days, the nuts were taken out from the refrigerator to calculate percentage of the nuts released from dormancy, and the nuts which were split their shells consider as released nut from dormancy. 5% Duncan's Multiple Range Test used for comparison of the means.

**Calculation of germination percentage**

After calculating percentage of the nuts released from dormancy, the nuts which were released from dormancy directly sown to calculate germination percentage in a medium of fine sand in polyethylene bags with a size of 10×30 cm. In each bag one nut was sown without application of any treatment or disinfectant to the released nut from dormancy. The bags laid out in RCBD with three replications, each replication contained five sown nuts. Overall, 15 nuts were sown for every genotype in the three replications. The genotypes which were not released enough nuts (15 nuts) from dormancy (their nut shells were not split), unreleased ones were sown to complete the number. The experiment was practiced in a lath house condition, and the temperature inside the lath house were shown in figure (2). On September, 19 2021, the germination percentage was calculated, and 5% Duncan's Multiple Range Test used for comparison of the means.

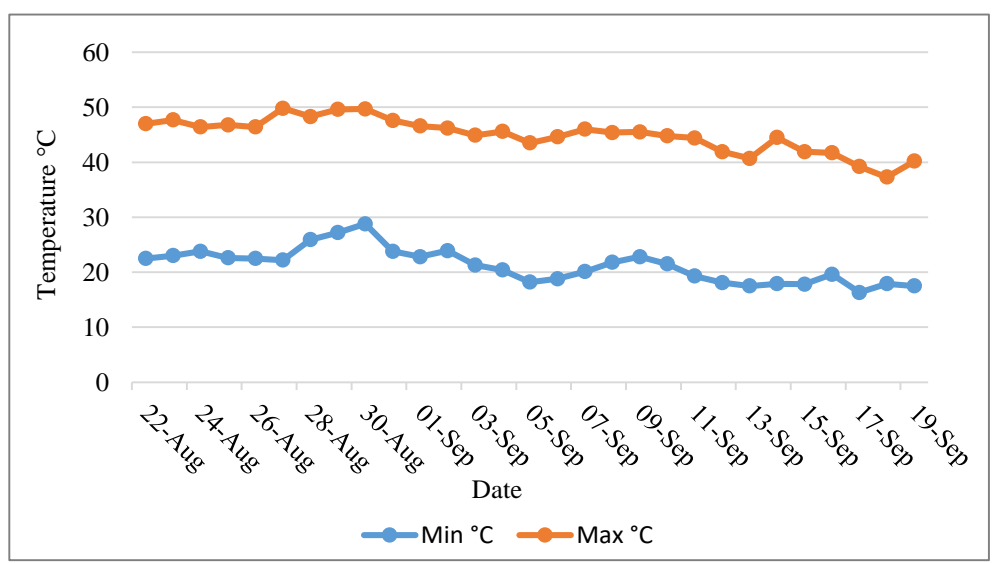

Figure (2) Daily minimum and maximum temperature (°C) inside the lath house started from August 22, 2021 to September, 19 2021

All data from nut trait measurements, estimation of percentage of the nuts released from dormancy and calculation of germination percentage were analyzed by using XLSTAT software version 2019.2.2, one-way ANOVA.

**RESULTS AND DISCUSSION**

The genotypes of bitter almond showed different nut measured traits (Table1). Nut length was high (32.73 mm) in the nuts of genotype 3 (G3), while genotype 7 (G7) and then genotype 4 (G4) demonstrated the best (21.84 and 21.25 mm, respectively) nut width. Additionally, nut size was increased to the best (4.26 and 4.17 cm$^3$) in the nuts of both genotype 8 (G8) and 4 (G4), respectively. Genotype 8 gave the maximum (4.13 g) nut weight, shell weight (3.27 g) and shell thickness (3.37 mm), and kernel weight reached the highest (0.99 g) in the nuts of genotype 5 (G5). In opposite, the lowest values of nut length (23.66 mm), nut width (18.77 mm), nut size (3.16 cm$^3$), shell weight (1.94 g) and shell thickness (2.31 mm) were found in the nuts of genotype 2 (G2). Moreover, nut weight (2.7 g) in the nuts of genotype 6 (G6) and kernel weight (0.6 g) in the nuts of genotype 1 (G1) were minimum values. The longest nut in genotype 3 may be due to that the nuts of this genotypes had a sharp apex and rendered them a long shape, but the apex was not found in the nuts of other genotypes with exception of the nuts of genotypes 9 (G9). Also, G9 gave the second longest nut length (29.58 mm). On the other hand, the results of the other traits might be because of interrelation the traits among themselves. Based on this, high or low values of nut width and size in the genotypes were in interdependence to each other, mostly the nuts statistically had high nut width value also had high nut size, and vice versa (figure 5). Besides, weight of nuts was mostly dependent on the shell thickness, the nuts of the both G8 and G7 demonstrated superior shell thickness and they had high nut weight as well. It is noteworthy to point out that no double kernel was found in the nuts examined to measure the traits, and just one empty nut was observed in G8. In this regard, other researcher found that almond genotypes crucially characterized the nuts. Imani and Nagoya (2006) found that nuts of almond genotypes were different in nut size, nut shape, shell color and percentage of double kernel. Sorkheh *et al*. (2010) evaluated 46 almond genotypes and they indicated that genotype was decisively determined all nut and kernel traits in a highly significant rate. Furthermore, Khadivi *et al* (2019) summarized that almond genotypes could be select for breeding depending on nut and kernel traits, and they showed that some traits of nuts and kernels correlated with each other. The variation in nut and kernel traits are under control of genetic and environmental controls. Sakar *et al*. (2019) found that genotypic and environmental effects equally regulated the traits of nut and kernel in almond cultivars; they also referred that nut and kernel length and sphericity were genetically regulated, but nut and kernel thickness were environmentally regulated.

**Table (1) Nut traits of the 10 genotypes of bitter almond**

| Genotypes | Nut length (mm) | Nut width (mm) | Nut size (cm$^3$) | Nut weight (g) | Kernel weight (g) | Shell weight (g) | Shell thickness (mm) |
|---|---|---|---|---|---|---|---|
| G1 | 28.85 bc | 19.89 cde | 3.76 ab | 2.78 c | 0.6 f | 2.18 c | 2.87 b |
| G2 | 23.66 f | 18.77 f | 3.16 c | 2.72 c | 0.767 cde | 1.94 c | 2.31 d |
| G3 | 32.73 a | 19.34 ef | 3.97 a | 3.61 b | 0.76 de | 2.85 b | 2.86 b |
| G4 | 28.13 cd | 21.25 ab | 4.17 a | 3.60 b | 0.90 b | 2.69 b | 2.79 b |
| G5 | 26.61 e | 20.94 abc | 3.76 ab | 3.52 b | 0.99 a | 2.52 b | 2.54 c |
| G6 | 27.55 de | 19.27 ef | 3.25 bc | 2.67 c | 0.69 e | 1.98 c | 2.37 d |
| G7 | 26.78 e | 21.84 a | 3.96 a | 3.66 b | 0.83 bcd | 2.83 b | 3.23 a |
| G8 | 28.03 cd | 20.83 abc | 4.26 a | 4.13 a | 0.85 bc | 3.27 a | 3.37 a |
| G9 | 29.58 b | 19.59 def | 3.777 ab | 3.38 b | 0.79 cd | 2.58 b | 2.81 b |
| G10 | 28.41 cd | 20.49 bcd | 3.773 ab | 3.63 b | 0.78 cde | 2.84 b | 2.74 b |

The genotypes share the same letter for the same trait do not differ significantly ($P \leq 0.05$) according to Duncan's Multiple Range Test

The data in figure (3) explained that the nuts of the 10 bitter almond genotypes differently responded to the stratification treatment at 6±1°C for 55 days in refrigerator. The nuts of G6 and G4 were very readily responded to release from dormancy, whereas an unsatisfactory response was found in nuts of G7 and G1 to release from dormancy. The highest number (100%) of nuts released from dormancy in G6 and to a lesser extent (86.11%) in the nuts of G4. In contrast no nuts were released from dormancy in G7, and G1 gave (11.63%) the nuts released from dormancy. Dormancy in almond nuts is restricted by a physical barrier arises from nut shell (endocarp) which is impermeable to water and gases, or physiological factors exist in embryo, such as ratio of abscisic acid (ABA) to gibberellic acid (GA) which are adjusted by environmental factors, especially temperature and light (Zeinalabedini *et al*., 2009). In almond nut shell plays significant role in release the nuts from dormancy. The nut shell of almond act as obstacle to gas exchange, imbibition and washing out of inhibitor substances. So, in the current study it was clear that the nuts of G6 had a thin shell and G7 had a high shell thickness (Table 1). The researcher found that the treatments carried out to reduce the negative effect of nut shell in almond positively released the nuts from dormancy and facilitated germination process. Rafiei *et al*. (2019) found that scarification of almond nut shell increased the capacity of the seeds to release from dormancy and germination. García-Gusano *et al*. (2005) observed that removing of nut shell in almond resulted in the same effect of stratification on the nuts with shell. Apart from the role of nut shell in almond nut dormancy, in the present study G8 gave (52.5%) the nuts released from dormancy, however they had the highest (3.37 mm) shell thickness (Table 1). This may be due to seed dormancy is not just related to shell thickness but other factors such as high inhibitors might influence seed dormancy; perhaps the level of the inhibitors were different according to the genotypes. In this connection, Al-Imam and Qrunfleh (2007) concluded that reduction in abscisic acid (ABA) which is an inhibitor observed in bitter almond nuts during stratification and associated with better release from dormancy and germination. Additionally, Mohammadi *et al*. (2014) referred that ABA content in almond nut varieties was variable, K-66 and Rabie almond varieties had the highest ABA concentration.

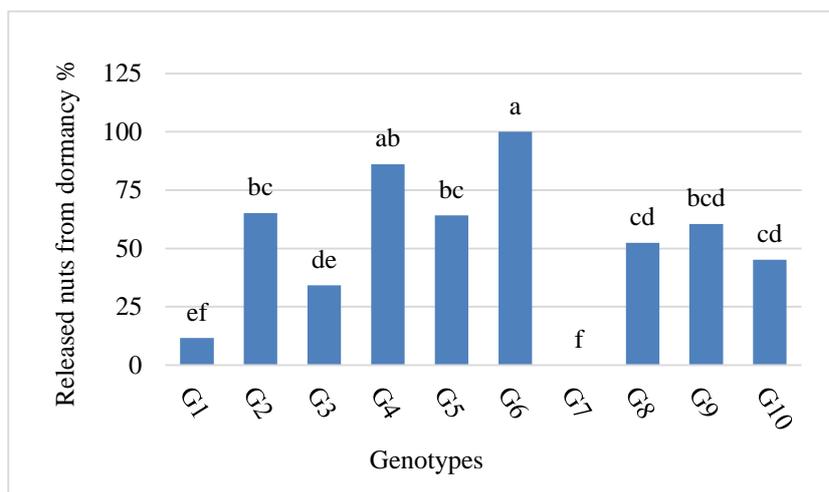

Figure (3) The percentage of the nuts of the 10 bitter almond genotypes released from dormancy after stratification at 6±1°C for 55 days. The genotypes share the same letter for the same trait do not differ significantly ($P \leq 0.05$) according to Duncan's Multiple Range Test

The figure (4) presented the results of germination percentage of the 10 bitter almond genotypes after sowing and waiting for 29 days, after release from dormancy. The data confirmed that the genotypes demonstrated different gernmination percentage; the G2 and G6 equally gave the best (73.33%) gemination followed by G4 (46.66%). Inversely, no germinated nuts were obtained in G7, and the lowest (6.66%) germination observed in th nuts of G1

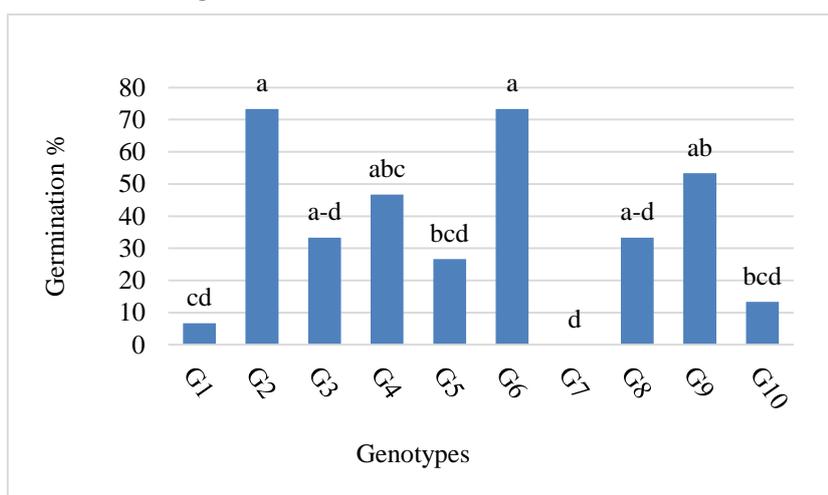

Figure (4) germination percentage of the nuts of the 10 bitter almond genotypes after 29 days. The genotypes share the same letter for the same trait do not differ significantly ($P \leq 0.05$) according to Duncan's Multiple Range Test

The differences in germination ability of the genotyps in the present study might belonge to the their differences in the ability to release from dormancy. The G6, G4 and G2, respectively were those genotypes released from dormancy in a high number, while G7 and G1 were the genotypes trifling released from dormancy. Release from dormancy is the first step for a successful germination preocess in the dormant seeds (Bentsink and Koornneef, 2008). It has been revealed in almond that germination is variable according to genotype. Taoueb *et al*. (2017) discovered that the nuts of almond genotypes differently responded to the same stratification period to improve germination percentage.

The person correlation matrix (Figure 5) made clear percentage of the nuts released from dormancy positively correlated with germination percentage (r = 0.83, p-value = 0.003). Germination percentage negatively associated with nut width (r = -0.65, p-value = 0.042) and shell thickness (r = -0.64, p-value = 0.047). Nut width positively linked to nut size (r = 0.7, p-value = 0.024) and nut weight (r = 0.67, p-value = 0.033). Moreover, nut size was in positive relationship with nut weight (r = 0.87, p-value = 0.001), shell weight (r = 0.89, p-value = 0.001) and shell thickness (r = 0.84, p-value = 0.002). Nut weight positively correlate with shell thickness (r = 0.75, p-value = 0.013). A positive link was noticed between shell weight and shell thickness (r = 0.83, p-value = 0.003).

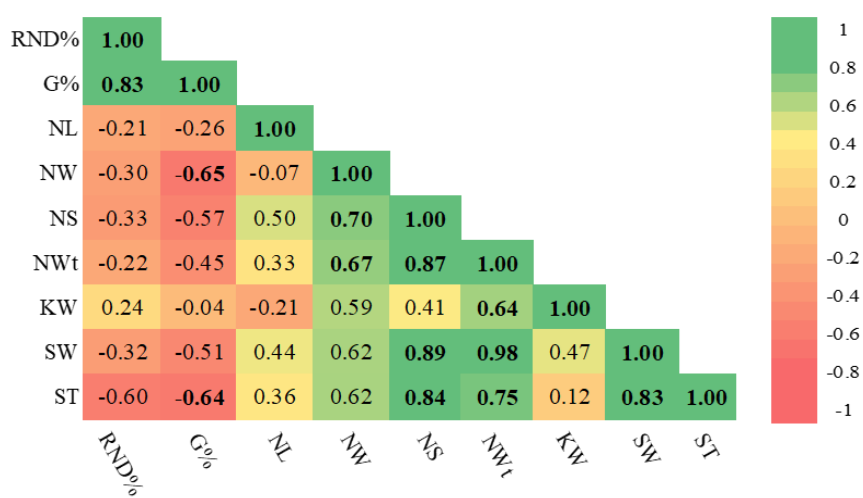

Figure (5) Person correlation of the traits. RND%: Released nuts from dormancy, G%: Germination%, NL: Nut length, NW: Nut width, NS: Nut Size NWt: Nut weight, KW: Kernel weight, SW: Shell weight, ST: Shell thickness

**CONCLUSION**

The results of released percentage of nuts from dormancy and germination percentage confirmed that G6, G4 and G2 gave the promising consequences. Also, correlation of the parameters showed germination percentage relied on breaking dormancy, and shell thickness negatively affected germination percentage. Additionally, shell thickness positively related positively to nut size and nut weight.

**فحص بعض السمات وكسر السكون والقدرة على الإنبات في بعض الانماط الوراثية للَّوز المر**

أنور محمد رؤوف محمود    كوجر عمر صالح    آرام أكرم محمد

قسم البستنة - كلية علوم الهندسة الزراعية - جامعة السليمانية

### الخلاصة

**الكلمات المفتاحية:** اللَّوز المر، صفات الجوزة (ثمرة اللَّوز)، السكون، الانبات.

اجريت هذه الدراسة في كلية علوم الهندسة الزراعية، جامعة السليمانية، اقليم كردستان العراق بهدف دراسة بعض صفات عشره انماط وراثية لثمرة اللَّوز المر، وقدرتها على كسر السكون والإنبات. فتم اخذ قياسات ثمار اللَّوز، ثم وضع الثمار في طبقات من الرمل وأدخلت في الثلاجة على درجة حرارة 6±1 لمدة 55 يوماً، ثم بعد ذلك تم زراعة البذور المنضدة في وسط غريني ناعم بتأريخ 22/آب/2021 لمدة 29 يوماً لاحتساب نسبة الإنبات. فسجلت اختلافات كبيرة بين انماط اللَّوز المزروعة في صفات الثمار، وتراوح طول الثمار بين (23.66-32.73مم)، وعرضها بين (18.77-21.84مم)، وحجمها كذلك تراوح بين (3.16-26.4 سم$^3$)، و وزنها بين (2.67-4.13غم)، و وزن النواة (0.6-0.99غم)، و وزن القشور تراوح بين (1.94-3.27غم)، وسمك القشرة (2.31-3.37مم). أظهرت النتائج في كسر السكون و نسبة الإنبات ان اعلى قيمة في كسر السكون كانت في الطرز الوراثية G6 (100%) و G4 (86.11%) و G2 (65.22%) ونفس الطرز اعطت اعلى نسبة مئوية الإنبات البذور، خاصة الطرازان الوراثيان G6 و G2 سجلا (73.33%) من نسبة الإنبات. اعتماداً على نتائج كسر السكون و النسبة المئوية للإنبات كان أفضل الطرز هما G6 و G2 بجانب طراز G4.